\documentclass[lettersize,journal]{IEEEtran}
\usepackage{amsmath,amsfonts,amssymb}
\usepackage[caption=false,font=normalsize,labelfont=sf,textfont=sf]{subfig}
\usepackage{graphicx}
\usepackage{cite}
\usepackage{xcolor}
\usepackage{tikz}
\usepackage{hyperref}
\usepackage[capitalise,nameinlink]{cleveref}
\usepackage{afterpage}
\usepackage{multirow} 
\usepackage{booktabs}

\usepackage[table]{xcolor}

\Crefname{equation}{Eq.}{Eqs.}
\Crefname{figure}{Fig.}{Figs.}
\Crefname{tabular}{Table}{Tables}
\Crefname{section}{Section}{Sections}

\def\BibTeX{{\rm B\kern-.05em{\sc i\kern-.025em b}\kern-.08em
    T\kern-.1667em\lower.7ex\hbox{E}\kern-.125emX}}

\definecolor{lime}{HTML}{A6CE39}
\DeclareRobustCommand{\orcidicon}{%
    \begin{tikzpicture}
    \draw[lime, fill=lime] (0,0) 
    circle [radius=0.16] 
    node[white] {{\fontfamily{qag}\selectfont \tiny ID}};
    \draw[white, fill=white] (-0.0625,0.095) 
    circle [radius=0.007];
    \end{tikzpicture}
    \hspace{-2mm}
}

\foreach \x in {A, ..., Z}{%
	\expandafter\xdef\csname orcid\x\endcsname{\noexpand\href{https://orcid.org/\csname orcidauthor\x\endcsname}{\noexpand\orcidicon}}
}

\makeatletter
\def\ps@IEEEtitlepagestyle{%
  \def\@oddfoot{\mycopyrightnotice}%
  \def\@oddhead{\hbox{}\@IEEEheaderstyle\leftmark\hfil\thepage}\relax
  \def\@evenhead{\@IEEEheaderstyle\thepage\hfil\leftmark\hbox{}}\relax
  \def\@evenfoot{}%
}
\def\mycopyrightnotice{%
  \begin{minipage}{\textwidth}
  \centering \scriptsize
  This work has been submitted to the IEEE for possible publication. Copyright may be transferred without notice, after which this version may no longer be accessible.
  \end{minipage}
}
\makeatother

\begin{document}
\title{Bio-Inspired Electrotactile Feedback: A Framework and Impacts of Multimodal Cues}

\author{
Melika~Emami\textsuperscript{§}\orcidA, Amirhossein~Bayat\textsuperscript{§}\orcidB, Rahim~Tafazolli\orcidC,~\IEEEmembership{Senior Member,~IEEE,}
and Atta~Quddus\orcidD%
\thanks{\textsuperscript{§}Equal contribution.}%
\thanks{All authors are with the Institute for Communication Systems (ICS), Home of 5G \& 6G Innovation Centres, University of Surrey, UK.}
}

\maketitle

\bstctlcite{IEEEexample:BSTcontrol}

\begin{abstract}
Electrotactile stimulation, as a haptic feedback method, often struggles to evoke high-quality natural sensations. Consequently, despite its practical advantages, it is generally less preferred in human-machine systems than mechanical approaches. In this paper, we propose a bio-inspired stimulation framework that generates signals modelled on afferent neural activity derived from a touch model. By producing more biologically plausible spiking patterns, the proposed approach aims to enhance the realism of perceived tactile sensations.
To evaluate these stimulation patterns in a realistic task context and to quantify the influence of visual and movement cues, we conducted a series of subjective experiments under three conditions. The results indicate that bio-inspired modulation was effective: in particular, bio-inspired co-modulation of amplitude and frequency achieved a favourable pressure-irritation trade-off, increasing the probability of perceiving pressure while maintaining low irritation (pressure: 0.78; irritation: 0.24) relative to the other patterns. Visual cues and hand movement further improved feedback quality by significantly increasing the probability of reporting pressure (p$<$0.001) without a corresponding increase in irritation, and by increasing mean naturalness ratings.
Overall, these findings demonstrate the potential of bio-inspired electrotactile patterns to improve haptic feedback quality, and highlight the importance of evaluating such methods under realistic, multimodal interaction scenarios.
\end{abstract}


\section{Introduction}
\IEEEPARstart{H}{aptic} feedback enables the perception of touch in human-machine systems such as virtual reality (VR), teleoperation, remote surgery, social interaction, and prosthetics \cite{pacchierotti2023cutaneous, raisamo2022interpersonal}. By conveying tactile sensations, haptics can enhance realism, immersion, and precision in these experiences.

Haptic technologies are commonly categorised as mechanical or electrical \cite{emami2024survey}. Mechanical approaches (e.g., piezoelectric and vibrotactile actuators) physically interact with the skin or joints to generate sensations such as touch, pressure, or grasping \cite{huang2021recent}. While often capable of natural sensations, they can be bulky, complex, power-hungry, and limited in spatial resolution. Electrotactile feedback instead stimulates peripheral nerves using low-current pulses delivered through skin electrodes \cite{saunders1983information, kajimoto2004electro}. It is lightweight, compact, flexible, energy efficient and can support high-resolution stimulation \cite{kourtesis2022electrotactile}; however, it frequently evokes artificial or ``electrical'' percepts that reduce perceived realism \cite{raspopovic2021sensory}.

Therefore, improving the quality of the electrotactile sensation is critical. Prior work has optimised parameters such as frequency in tonic stimulation to shape percepts and reduce undesirable sensations (e.g., tingling) \cite{yang2019electrotactile, manoharan2024characterization, zhou2022electrotactile}. Beyond single-parameter tuning, spatio-temporal patterns for signal modulation or electrode activation have been proposed to better reflect the dynamics of real touch \cite{paredes2015impact}. For example, linear amplitude modulation has been used to convey softness by mapping stimulus intensity to virtual interaction forces \cite{yem2018effect}, and modulating the effective contact area via spatio-temporal stimulation has also been shown to influence softness perception \cite{suga2023softness}. Although such approaches exploit the physics of interaction, they may still fail to reproduce the underlying neural responses that accompany natural touch.

A complementary direction is to approximate biological neural coding of touch using bio-inspired modulation strategies \cite{saal2015biomimetic}. In prosthetics, biomimetic sensory feedback has produced more intuitive and lifelike sensations than conventional approaches \cite{saal2015biomimetic, formento2020biomimetic, okorokova2018biomimetic, greenspon2023biomimetic}. However, existing frameworks raise two key issues for non-prosthetic contexts. First, most evidence comes from prosthetic applications, where stimulation site, modality, and objectives differ from VR and teleoperation; consequently, the appropriate bio-inspired design may not directly transfer. Second, evaluation methods should match application goals: prosthetics prioritise sensory restoration and functional integration, whereas VR and teleoperation often emphasise immersion, naturalness, and perceptual precision.

A further limitation is ecological validity. To stabilise skin impedance and electrode contact, many electrotactile studies require a stationary hand \cite{garenfeld2023novel}. Although this improves reliability, it does not reflect dynamic scenarios in which movement is integral, such as VR interactions. Task-based studies that include hand gestures often focus on task performance rather than perceived sensation quality \cite{yoshimoto2015electrotactile}.

In this study, we pursue three objectives: (1) propose a framework for generating bio-inspired electrotactile stimulation patterns to improve perceived sensation quality; (2) evaluate these patterns under more realistic, application-relevant conditions (VR-like interaction); and (3) quantify how visual and movement cues shape the haptic experience.

Our framework uses TouchSim, a computational touch model that simulates afferent activity in response to skin deformation \cite{saal2017simulating}. By aggregating model outputs (e.g., firing rates), we map biologically derived signals to electrotactile parameters, including amplitude and frequency. We then conducted a three-session subjective study: (i) no additional cues, (ii) synchronised video of a virtual hand pressing a button (visual cues), and (iii) participants mimicking the button press in mid-air while viewing the same video (visual + movement cues). Sensation quality was assessed via reported sensation types (pressure and irritation) and perceived naturalness.

Overall, the results show that bio-inspired patterns can enhance electrotactile sensation quality by increasing pressure while keeping irritation comparatively low, with the most favourable balance achieved by hybrid co-modulation of amplitude and frequency derived from the proposed framework. Moreover, synchronised video and movement cues further improved perceived quality for most patterns, with particularly strong gains for bio-inspired signals, suggesting that combining bio-inspired stimulation with multimodal cues yields the most favourable outcomes.

The remainder of this paper is organised as follows. \Cref{sec::framework} describes the proposed framework. \Cref{sec::experimental_setup} details the stimulation signals, device, experimental procedure, and participants. \Cref{sec::results} reports the analysis and results, and \Cref{sec::discussion} discusses key findings, limitations, and future directions. Finally, \Cref{sec::conclusion} concludes the paper.

\section{Bio-Inspired Framework}
\label{sec::framework}
The bio-inspired framework first uses a touch model to simulate afferent responses to a predefined tactile interaction, and then maps these responses to electrotactile stimulation parameters. The two components are described below.

\subsection{Computing Afferent Responses}
To compute afferent responses to spatio-temporal mechanical stimuli, TouchSim is used, which is a computational model of tactile afferents innervating the palmar surface of the hand \cite{saal2017simulating}. It combines a mechanical model of skin deformation with a neural model that predicts spiking activity at the level of individual afferents, including rapidly adapting (RA), slowly adapting type 1 (SA1), and Pacinian corpuscle (PC) fibres.

Hence, we first designed a mechanical stimulus intended to evoke a pressure-like percept, which was our primary objective. We selected a button-press interaction because it is a common tactile event, yields a simple and repeatable gesture, and supports integration with a VR video for multimodal experimental conditions. A circular contact area centred on the index fingertip was defined, and pressure was applied to reach a maximum indentation of 3 mm. The stimulus followed a 3 s ramp-and-hold profile, with each ramp phase lasting 0.7 s. To emulate small vibrations associated with button pressing, low-level noise was added with a maximum amplitude of 0.02\% of the main stimulus amplitude.

\begin{figure}[t!]
    \centering
    \includegraphics[width=0.9\linewidth]{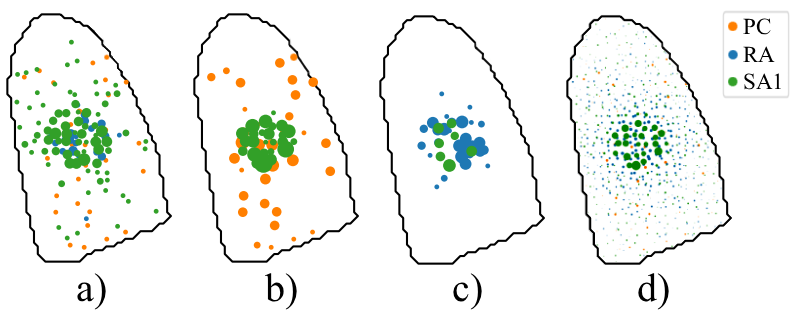}
    \caption{TouchSim responses to the mechanical stimulus, shown in 200 ms time bins at a) onset, b) mid-point, c) end, and d) whole duration of stimulation. Dot size indicates the relative total spike count.}
    \label{fig::touchsim_response}
\end{figure}

The computed afferent responses to the designed stimulus provide high-resolution spatial and temporal information regarding action potentials for each afferent. However, as directly reproducing individual spikes via surface electrotactile stimulation is impractical, this neural activity must be aggregated to extract actionable control parameters. The resulting aggregated action potentials, visualised over 200 ms windows at the onset, mid-point, end of stimulation, and also over the whole duration, are presented in \Cref{fig::touchsim_response}.

Building upon prior approaches that used metrics such as activation area and firing counts, we characterise afferent responses using three features aggregated within time bins: (i) active afferent location, (ii) firing rate, and (iii) activation rate (the proportion of active afferents) \cite{okorokova2018biomimetic, formento2020biomimetic}. These features capture the core spatio-temporal structure of the neural response, forming a robust basis for mapping electrotactile parameters to biologically derived patterns. We incorporate all three afferent classes (RA, SA1, and PC) into this framework, consistent with findings that combining encodings across afferent types yields more natural percepts \cite{valle2018biomimetic}.

To select an appropriate bin duration, we evaluated multiple bin sizes and found that 50 ms provided the best trade-off. Larger bins excessively smoothed temporal structure, whereas smaller bins introduced noise and increased computational demand, consistent with prior observations \cite{valle2024biomimetic}. 

\subsection{Relating Afferent Responses to Electrotactile Parameters}

To map the aggregated afferent responses to electrotactile parameters, the proposed framework was developed in a structured manner, beginning with the active afferent locations and then incorporating firing rate and activation rate features.

\subsubsection{Location to Electrode}
\label{subsub::location_to_electrode}
The perceived location of electrotactile sensations, determined by the spatial pattern in which most afferent fibres are recruited, is largely governed by electrode geometry and placement, because current density is highest beneath the electrode pads \cite{isakovic2022impact, patel2016multichannel}. In principle, TouchSim provides the locations of active afferents in each time bin, which could be clustered and targeted using multiple electrode pads. Each pad could then deliver either a common carrier signal (e.g., tonic 50 Hz) or a cluster-specific waveform.

However, because electrode geometry is fixed over time, recruiting time-varying spatial clusters of afferents would require either (i) a general multi-pad array with pads positioned near approximate cluster locations across bins, or (ii) a custom electrode layout tailored to the dominant spatial activation observed in the simulated responses. We elaborate the chosen method in this study in \cref{sub::electrode_design}.

\begin{figure}[t!]
    \centering
    \includegraphics[width=\linewidth]{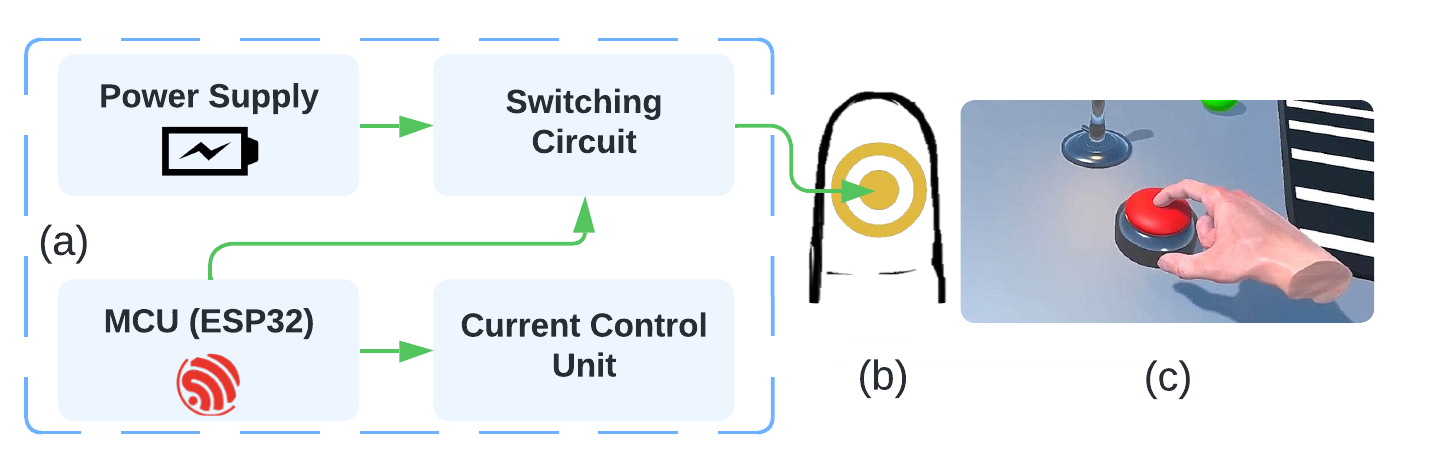}
    \caption{Experimental setup. (a) Electrotactile device capable of generating bio-inspired stimulation signals. (b) Electrode geometry and placement on the fingertip.}
    \label{fig::setup_diagram}
\end{figure}

\subsubsection{Firing Rate to Frequency} 
Spatial grouping based on the locations of active afferents, generally, yields a set of spatial clusters. For cluster $s$ in time bin $t$, the firing rate $FR_{t,s}$ was computed by normalising the total number of action potentials by the number of active afferents in that cluster and the bin duration:

\begin{equation}
FR_{t,s} = \frac{\sum_{i=1}^{K_{t,s}} AP_i}{K_{t,s} \, T_t},
\label{eq::bin_firing_rate}
\end{equation}
where $AP_i$ is the number of action potentials generated by afferent $i$ within bin $t$ and cluster $s$, $K_{t,s}$ is the number of active afferents in that bin and cluster, and $T_t$ is the bin duration (s). This yields a time-normalised measure of firing activity for each cluster while accounting for changes in afferent recruitment.

Stimulation frequency ($Freq$) primarily governs how often recruited neurons are driven and therefore strongly influences firing rate. Pulse width ($PW$) can also contribute if it is sufficiently long to elicit multiple spikes per pulse \cite{formento2020biomimetic, graczyk2016neural}. Accordingly, for each bin and cluster we define
\begin{equation}
f_{t,s}\!\left(Freq_{t,s}, PW_{t,s}\right)=FR_{t,s},
\label{eq::our_firing_rate_formula}
\end{equation}
where $f_{t,s}$ relates the stimulation parameters delivered by a single electrode pad in a fixed time bin to the firing rate of afferents in the corresponding cluster.

To isolate the role of frequency and reduce experimental complexity, we fixed $PW$ at a value intended to elicit at most one action potential per afferent per pulse. This improves interpretability and reduces calibration burden, since $PW$ also affects perceived intensity. We fixed $PW$ rather than $Freq$ because frequency is a key determinant of electrotactile quality and central to reproducing bio-inspired temporal structure \cite{zhou2022electrotactile, graczyk2022frequency, kajimoto1999tactile}. Assuming a linear mapping under fixed $PW$, we obtain
\begin{equation}
Freq_{t,s} = A_1 FR_{t,s} + B_1,
\label{eq::linear_freq}
\end{equation}
where $A_1$ and $B_1$ are mapping parameters. 

From the TouchSim outputs, across time bins and clusters, $FR$ ranged from 1.0 to 33.33 (mean: 27.07). These values fall within frequency ranges commonly used in electrotactile stimulation \cite{graczyk2022frequency, emami2024survey}. Accordingly, setting $A_1=1$ and $B_1=0$ in \Cref{eq::linear_freq} maps $FR$ directly to stimulation frequency. This mapping is a pragmatic baseline and will be further tuned when defining the experimental stimulation signals (\Cref{sec::experimental_setup}).

\subsubsection{Activation Rate to Amplitude} 
The activation rate ($AR$) was defined as the proportion of active afferents relative to the total number of afferents within the targeted region. For cluster $s$ in time bin $t$, this was computed as
\begin{equation}
AR_{t,s} = \frac{K_{t,s}}{N_s},
\label{eq::activation_rate}
\end{equation}
where $K_{t,s}$ is the number of active afferents in bin $t$ and cluster $s$, and $N_s$ is the total number of afferents (active and inactive) within the spatial region associated with cluster $s$. To assign inactive afferents to clusters, each afferent was associated with the nearest cluster centroid (minimum Euclidean distance), thereby defining $N_s$.

In electrotactile stimulation, activation rate depends on electrode properties, stimulation amplitude, and pulse width. For example, increasing the active-pad area can enlarge the effective stimulation region and thereby recruit more afferents (given sufficient current density), or higher amplitude increases current spread and typically recruits more afferents \cite{formento2020biomimetic}. Pulse width can have a similar effect: because skin has non-zero permittivity, the transcutaneous potential rises approximately exponentially during a square pulse, and short pulses (e.g., $<5\tau$) may not reach peak potential, limiting recruitment. Accordingly, we define
\begin{equation}
g_{t,s}\!\left(Amp_{t,s}, Elec_s, PW_{t,s}\right)=AR_{t,s},
\label{eq::our_activation_rate_formula}
\end{equation}
where $g_{t,s}$ relates the stimulation parameters delivered by a single electrode pad to the activation rate of afferents in cluster $s$ within bin $t$.

In this study, electrode properties ($Elec_s$) was determined from the spatial distribution of active afferents, consistent with our sequential framework. Pulse width was also fixed to simplify stimulation and isolate frequency-modulation effects. Assuming a linear mapping under fixed $PW$ and $Elec_s$, from \cref{eq::our_activation_rate_formula} we obtain
\begin{equation}
Amp_{t,s} = A_2 AR_{t,s} + B_2,
\label{eq::linear_amp}
\end{equation}
where $A_2$ and $B_2$ are mapping parameters. 

From the TouchSim outputs, across time bins and clusters, the activation rate $AR$ had low magnitude and high variability (min: 0.001, max: 0.068, mean: 0.040), which is generally insufficient to reliably elicit sensation in most participants \cite{jin2024advances}. We therefore applied a modified min-max scaling to preserve temporal structure while expanding the usable amplitude range. The amplitude mapping parameters were computed as
\begin{subequations}
    \begin{align}
        A_2 &= \frac{\alpha}{\max(AR) - P_5(AR)},
        \label{eq::scaling_factor_formula::a}\\
        B_2 &= -A_2 \cdot \max(AR) + \beta,
        \label{eq::bias_formula::b}
    \end{align}
    \label{eq::scaling_and_bias_formula}
\end{subequations}
where $\alpha$ controls the modulation range and $\beta$ sets the peak amplitude. Instead of $\min(AR)$ (an outlier occurring at the end of stimulation), we used the 5th percentile $P_5(AR)$ to obtain a balanced scaling range while preserving the temporal pattern.

\section{Experimental Setup}
\label{sec::experimental_setup}
This section describes the experimental setup, including the designed electrodes and stimulation signals, device and interfaces, experimental procedure, and participant details.

\subsection{Electrode Design}
\label{sub::electrode_design}
Upon examination of the spatial response patterns seen in \cref{fig::touchsim_response}, in the time bins, the activity was concentrated within the mechanically stimulated region, with limited spread beyond it. Therefore, for the electrode pad design, we considered the dominant activation region (second approach mentioned in \cref{subsub::location_to_electrode}) and implemented a targeted configuration with a small active pad centred within the dominant activation region and a surrounding reference pad placed around the fingertip periphery (\Cref{fig::setup_diagram}b). This geometry encourages current flow outward from the fingertip centre, promoting activation beneath the intended site while limiting recruitment in surrounding regions. The reference pad area was approximately 5.3$\times$ larger than the active pad, reducing current density and mitigating unwanted sensations \cite{isakovic2022impact}.

We hypothesise that this configuration preserves a centralised percept while allowing only sparse, low-level activation outside the target area, consistent with the TouchSim spatial profile. Furthermore, adding additional pads would substantially increase calibration and experimental complexity, particularly when amplitude is also modulated.

\subsection{Stimulation Signals}
To examine how bio-inspired modulation influences perceived sensation quality in electrotactile feedback, we designed two sets of stimulation signals, as described below.

\subsubsection{Bio-inspired Modulation Set}
To isolate the effects of modulation, we implemented three strategies (frequency-only, amplitude-only, and co-modulation), each with a low-frequency (LF) and high-frequency (HF) variant:

\begin{itemize}

    \item \textbf{Bio frequency modulation:} frequency followed \Cref{eq::linear_freq} and amplitude was fixed at the participant's calibrated level. A second frequency-modulated signal was created by scaling frequency such that the mean increased from 27.07~Hz to 100~Hz, using $A_1=\frac{100}{\mathrm{mean}(FR)}$.
    
    \item \textbf{Bio amplitude modulation:} amplitude followed \Cref{eq::linear_amp} with carrier frequency fixed at either 20~Hz (consistent with the unscaled $FR$ range) or 100~Hz (consistent with the scaled range).
    
    \item \textbf{Bio hybrid modulation:} both amplitude and frequency were modulated simultaneously via \Cref{eq::linear_amp} and \Cref{eq::linear_freq}, using the same parameter mappings.
\end{itemize}

This yielded six bio-inspired signals: three in the LF regime and three in the HF regime. For frequency scaling, we used a multiplicative factor ($A_1$) rather than a bias term ($B_1$) to preserve relative temporal variations.

For amplitude modulation, based on pilot testing, $\alpha=0.3$~mA was set in \cref{eq::scaling_and_bias_formula}, which provided an effective compromise between preserving temporal variation and avoiding excessive fluctuations.

Because amplitude directly affects perceived intensity and discomfort, $\beta$ was participant-specific: we set $\beta$ such that each participant's peak amplitude lay between 0.5 and 3 mA, selected during calibration. This ensured that the waveform fluctuated within each participant's comfortable range, varying between their chosen peak and approximately 0.3 mA below it, following the temporal profile of $AR$.

The selected frequencies are based on the mechanoreceptors sensitives. The unscaled mean $FR$ (27~Hz) was obtained by temporal aggregation across all afferent types. Because mechanoreceptor sensitivities differ (RA1: $\sim$10-100~Hz, peak near 20~Hz; RA2: $\sim$60-800~Hz, peak near 250-300~Hz) \cite{emami2024survey, yang2019electrotactile, purves2019neurosciences}, we selected a target mean of 100~Hz for the scaled signals to probe the RA1-RA2 transition region. This choice also reflects common practice and supports perceptual separability while maintaining comfort \cite{kaczmarek2017interaction, manoharan2024characterization}.

\subsubsection{Reference stimuli}
To benchmark bio-inspired modulation against conventional designs, six additional reference signals were included:
\begin{itemize}
    \item \textbf{Linear amplitude modulation:} ramp-and-hold amplitude profile with fixed frequency at 20~Hz or 100~Hz.
    \item \textbf{Linear frequency modulation:} ramp-and-hold frequency profile from 20~Hz to 100~Hz with constant amplitude.
    \item \textbf{Hybrid linear modulation:} simultaneous amplitude and frequency ramp-and-hold modulation.
    \item \textbf{Tonic:} constant-parameter stimulation at 20~Hz and at 100~Hz.
\end{itemize}
These references enabled direct comparison between bio-inspired modulations, simple linear modulations, and non-modulated (tonic) stimulation.

In total, twelve signals were defined: BioAmp-LF, BioAmp-HF, BioFreq-LF, BioFreq-HF, CoBio-LF, CoBio-HF, LinAmp-LF, LinAmp-HF, LinFreq, CoLin, Tonic-LF, and Tonic-HF (``Co'' denotes co-modulation; LF/HF denote low/high frequency).

All signals used symmetric biphasic pulses with equal positive and negative phases of 300~\textmu s (total pulse duration: 600~\textmu s). Pulse width was chosen to be short enough to reduce the likelihood of multiple action potentials per pulse, while still allowing effective intensity calibration within device limits. In pilot testing (150-500~\textmu s), 300~\textmu s provided the best performance and was selected. Stimulus duration was 3~s, matching the mechanical stimulus. Biphasic pulses were used to maintain charge balance and reduce the risk of tissue damage, improving comfort \cite{emami2024survey}.

\subsection{Device and Interfaces}
Stimulation signals were delivered using the electrotactile device reported in \cite{bayat2025effects}, with the software adapted to support the proposed bio-inspired modulation patterns. The electrode geometry and placement followed the location-guided design described in the framework section. Electrodes were secured using a bandage, with tightness adjusted to participant comfort. To approximate practical use, electrodes were applied dry (i.e., without conductive gel).

Two graphical user interfaces (GUIs) were developed for calibration and evaluation to automate stimulus presentation, maintain double blinding, streamline data capture, and allow participants to proceed at their own pace.

\subsection{Experimental Procedure}
\subsubsection{Pre-experiment - Calibration}
The experiment began with a calibration phase to (i) avoid unpleasant sensations and (ii) standardise perceived intensity across stimulation patterns, thereby reducing confounding effects of intensity on subsequent ratings. Each pattern was presented at 26 intensity levels, with peak amplitude ranging from 0.5 mA to 3 mA. Participants rested their hand on a table and used the GUI to increment the amplitude from the lowest level until the sensation was clearly perceived and comfortable; this level was recorded as the optimal intensity. Participants calibrated each pattern individually and could revisit and compare previously selected levels to refine their choices.

\subsubsection{Exp.1 - Normal Mode}
After calibration, the first session evaluated the proposed patterns without additional cues, consistent with standard electrotactile protocols. To minimise bias, each stimulation pattern was presented three times in a randomised order, and participants were informed only of the total number of stimulations, yielding 36 trials. Participants maintained a resting hand position on the table to align with prior work \cite{garenfeld2023novel, bayat2025effects, isakovic2022impact}.

After each trial, participants selected one or more perceived sensation types from \emph{Pressure}, \emph{Irritation}, and \emph{No Sensation}. Irritation was described to participants as tingling or prickling. 
Naturalness was rated on a 6-point scale (0-5), where 0 indicated a completely unnatural sensation and 5 a completely natural sensation, consistent with prior studies \cite{bayat2025effects, valle2024biomimetic}. Here, naturalness was defined as the closeness of the perceived sensation to a real tactile experience, specifically the pressure sensation associated with pressing a button.

\subsubsection{Exp.2 - Video Mode}
In the second session, participants viewed a 3 s video of a virtual hand approaching and pressing a button, synchronised with the electrotactile stimulation (\Cref{fig::setup_diagram}c). Participants kept their hand resting on the table throughout each trial. The evaluation procedure matched Exp.1; however, to reduce session duration, each stimulation pattern was presented once, in random order.

\subsubsection{Exp.3 - Video+Movement Mode}
In the third session, participants synchronised their hand movement with the video by mimicking the button-press action. A full VR setup to capture hand motion was not available; therefore, participants were given sufficient practice time to align their movement timing with the virtual hand before rating sensations. The evaluation procedure remained the same as in the previous sessions.

Although calibration was conducted with the hand resting, prior evidence suggests that sensation thresholds are stable across hand postures \cite{yao2022encoding}. Accordingly, the calibrated intensity levels remained applicable for this movement condition.

\subsection{Participants}
Sixteen healthy participants (13 male, 3 female; mean age: 29.5~years) volunteered for the study. The purpose and procedure were provided in advance and reiterated on the day of testing. All participants gave written informed consent prior to participation. The study was approved by the University of Surrey Ethics Committee, UK (FEPS~23-24~019~EGA).

\section{Data Analysis and Results} \label{sec::results}
In this study, the primary objective was to improve sensation quality using a bio-inspired framework. In addition, by introducing video and movement cues, we examined how contextual information influences perceived sensation, both overall and in a pattern-specific manner. This section analyses the subjective outcomes using descriptive summaries and inferential modelling.

For inferential analysis, we used mixed-effects models implemented in \textsf{R}. These models were selected because data comprise repeated measurements (each stimulation was rated five times per participant), are unbalanced (trials removed following \emph{no sensation} reports), and do not satisfy assumptions required for traditional repeated-measures approaches (e.g., normality and sphericity). Moreover, mixed-effects models allow participant-specific random intercepts to capture stable between-participant differences in response tendency (e.g., consistently higher/lower reporting), improving the robustness of fixed-effect estimates.

\subsection{Impacts of Stimulation Patterns on Perceived Pressure}
As described above, participants reported the perceived sensation types for each trial, with \emph{Pressure} being the desired percept.

To model pressure, responses were dichotomised as present (1) versus absent (0), and a generalised linear mixed-effects model (GLMM) with a binomial error distribution was fitted. The stimulation pattern was included as a fixed effect and the participant as a random intercept. Separate models were fitted for each sensation-condition combination. For a sensitivity check, a single model was also fitted on the full dataset using the interaction term (i.e., \textit{Stimulation} * \textit{Condition}), which produced almost the same results. All models converged without errors. The estimated random-intercept variance indicated substantial between-participant heterogeneity in baseline pressure reporting, supporting the inclusion of participant-specific baselines.

Overall effects of stimulation pattern were evaluated using a likelihood-ratio test (LRT), comparing the full model (with stimulation pattern) against a random-intercept-only model. The stimulation pattern significantly improved model fit in the Normal condition ($\chi^{2}(11)=29.869$, $p<0.01$) and the Video condition ($\chi^{2}(11)=21.665$, $p<0.05$), but not in the Video+Movement condition ($\chi^{2}(11)=15.566$, $p>0.05$). Thus, under Video+Movement there was no clear evidence that pressure probability differed systematically across patterns after accounting for between-participant variability. Accordingly, pattern-specific post hoc comparisons were not conducted for Video+Movement and results for that condition are reported descriptively as exploratory.

Next, we obtained model-based predicted probabilities of reporting pressure for each stimulation pattern, together with 95\% confidence intervals. For inference, fixed-effect estimates were expressed as odds ratios relative to a chosen reference stimulation, with associated $p$-values.

Post hoc comparisons used treatment-versus-control contrasts (11 comparisons) to test each pattern against the reference while controlling the family-wise Type~I error rate; Holm correction was applied to the resulting $p$-values.

\begin{figure}[t!]
\centering
    \subfloat[Normal Condition]{\includegraphics[width=0.9\linewidth]{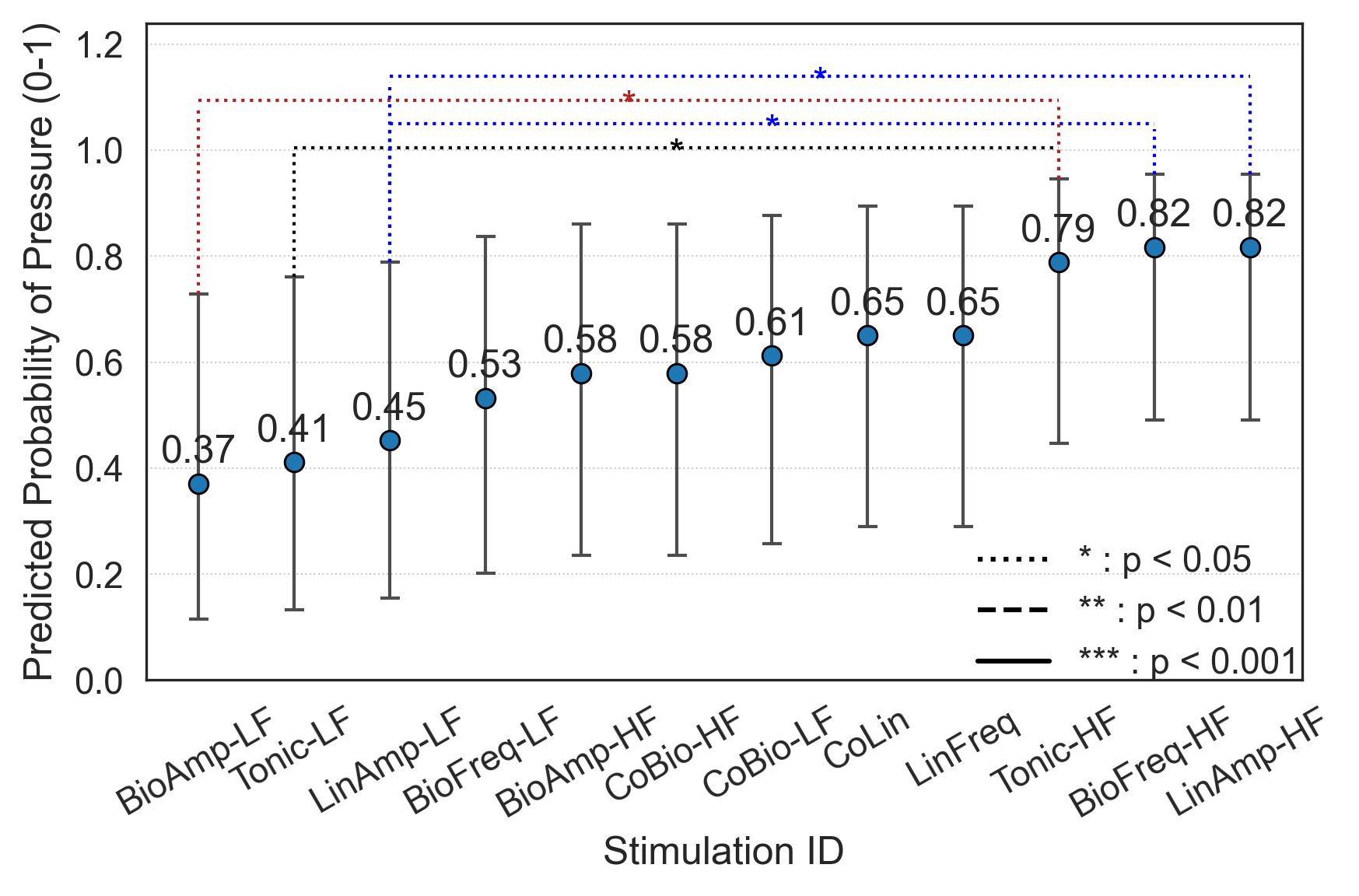}
    \label{fig::pressure_normal_probabilites}} 
    
    \subfloat[Video Condition]{\includegraphics[width=0.9\linewidth]{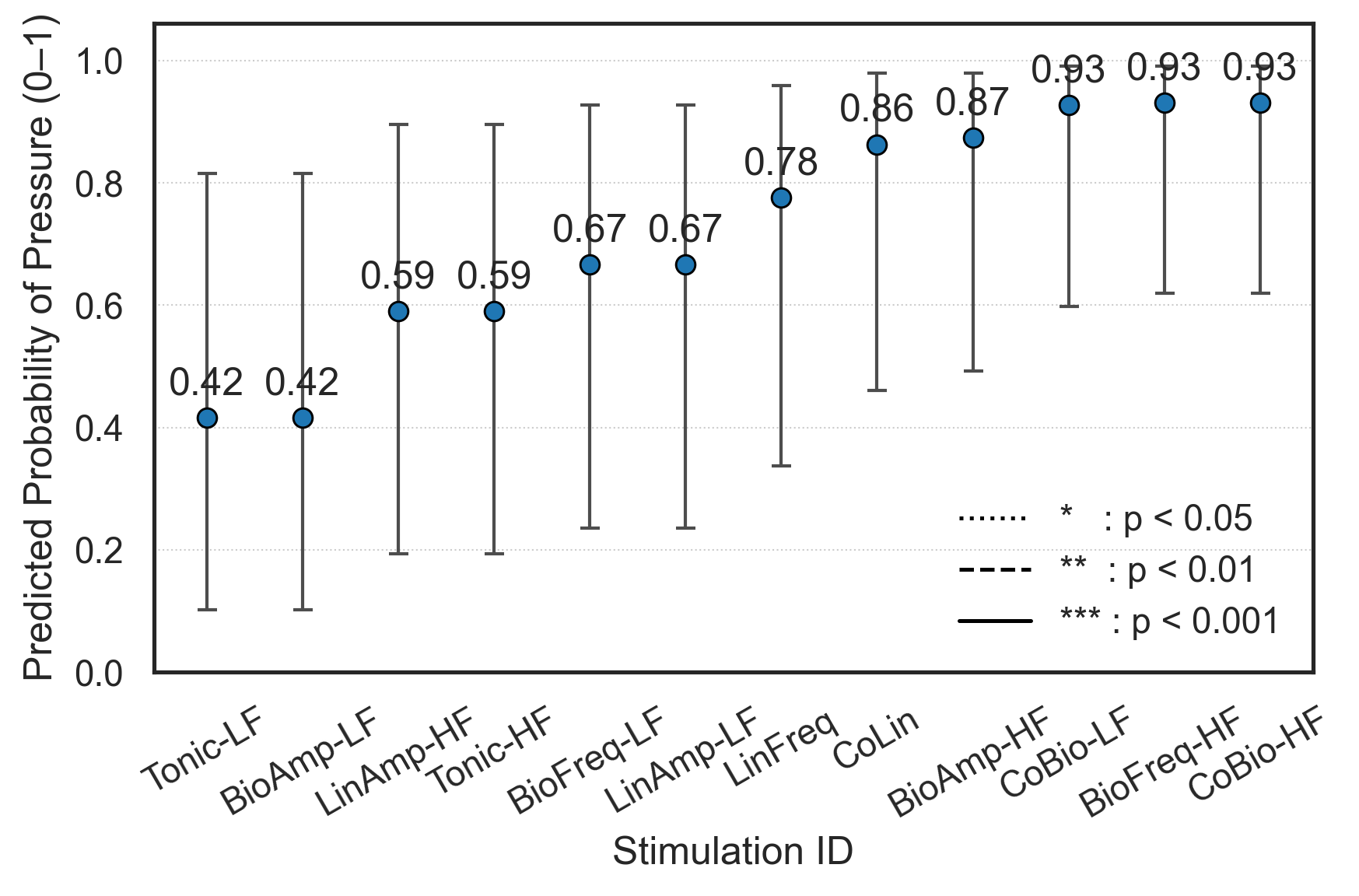}
    \label{fig::pressure_video_probabilites}} 
    
    \subfloat[Video+Movement Condition]{\includegraphics[width=0.9\linewidth]{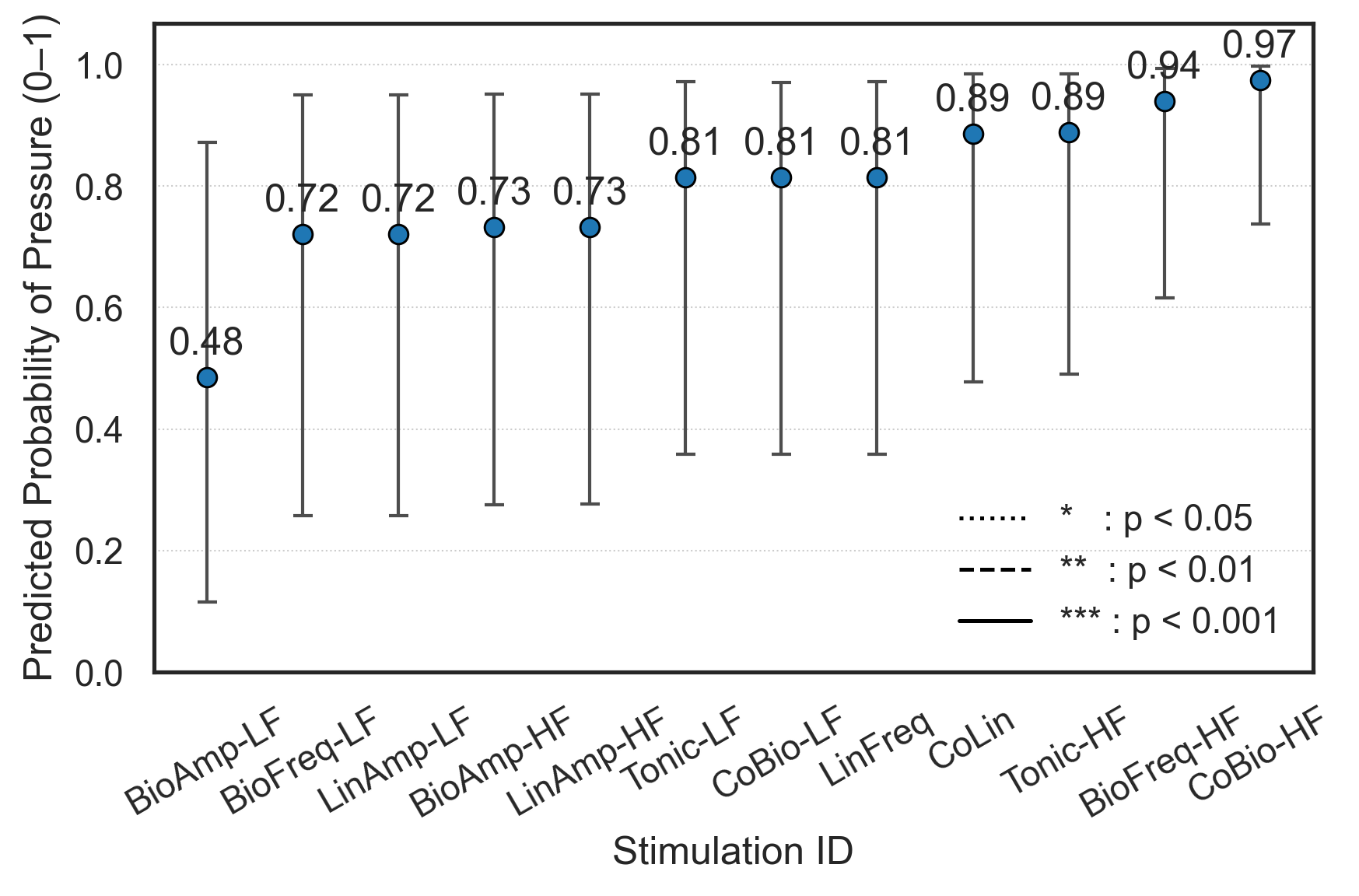}
    \label{fig::pressure_move_probabilites}} 
    
    \caption{The probability of producing pressure by each stimulation pattern under three different experimental conditions: a) Normal, b) Video, c) Video+Movement. Brackets with asterisks indicate the significant differences.}
    \label{fig::pressure_separate_condition}
\end{figure}

\Cref{fig::pressure_separate_condition} shows the stimulation patterns, ranked by predicted pressure probability, for each condition. To maintain visual clarity, we have omitted some redundant significance markers; brackets indicate only the nearest stimulation to the reference that reached statistical significance. Our analysis confirmed that all stimulations with a larger probability difference than this threshold (e.g., LinAmp-HF/BioFreq-HF vs. Tonic-LF/BioAmp-LF) were also significantly different from the reference.

In the Normal condition (\Cref{fig::pressure_normal_probabilites}), relative to Tonic-LF, Tonic-HF, BioFreq-HF, and LinAmp-HF showed significantly higher probabilities of reporting pressure ($p<0.05$). Using Tonic-HF as the reference, only BioAmp-LF and Tonic-LF were significantly lower. When LinAmp-HF (the highest-probability pattern) was used as the reference, none of the patterns ranked above LinAmp-LF differed significantly from LinAmp-HF.

In the Video condition, no contrasts remained significant after Holm correction. Before adjustment, and using Tonic-LF (the lowest-probability pattern) as the reference, CoLin, BioAmp-HF, CoBio-LF, BioFreq-HF, and CoBio-HF showed higher pressure probabilities; BioFreq-HF and CoBio-HF also differed from Tonic-HF prior to correction.

In the Video+Movement condition, the LRT indicated no overall effect of stimulation pattern. Accordingly, post hoc comparisons were not performed, and pattern-specific probabilities are reported descriptively as exploratory results.

\subsection{Impacts of Stimulation Patterns on Perceived Irritation}
The same modelling approach and inferential procedure were applied to the irritation sensation. The model fit was assessed using a LRT, comparing the full model (including stimulation pattern as a fixed effect) against a random-intercept-only model. The results (Normal: $\chi^{2}(11)=63.674$, $p<0.001$; Video: $\chi^{2}(11)=39.319$, $p<0.001$; Video+Movement: $\chi^{2}(11)=17.671$, $p>0.05$) indicated that stimulation patterns improved model fit in the Normal and Video conditions, but not in the Video+Movement condition. Therefore, under Video+Movement there was no clear evidence that the probability of perceiving irritation differed systematically across stimulation patterns after accounting for between-participant variability. Accordingly, pattern-specific post hoc comparisons were not pursued for Video+Movement, and any observed differences in that condition should be interpreted as descriptive/exploratory rather than confirmatory.

\begin{figure}[t!]
\centering
    \subfloat[Normal Condition]{\includegraphics[width=0.9\linewidth]{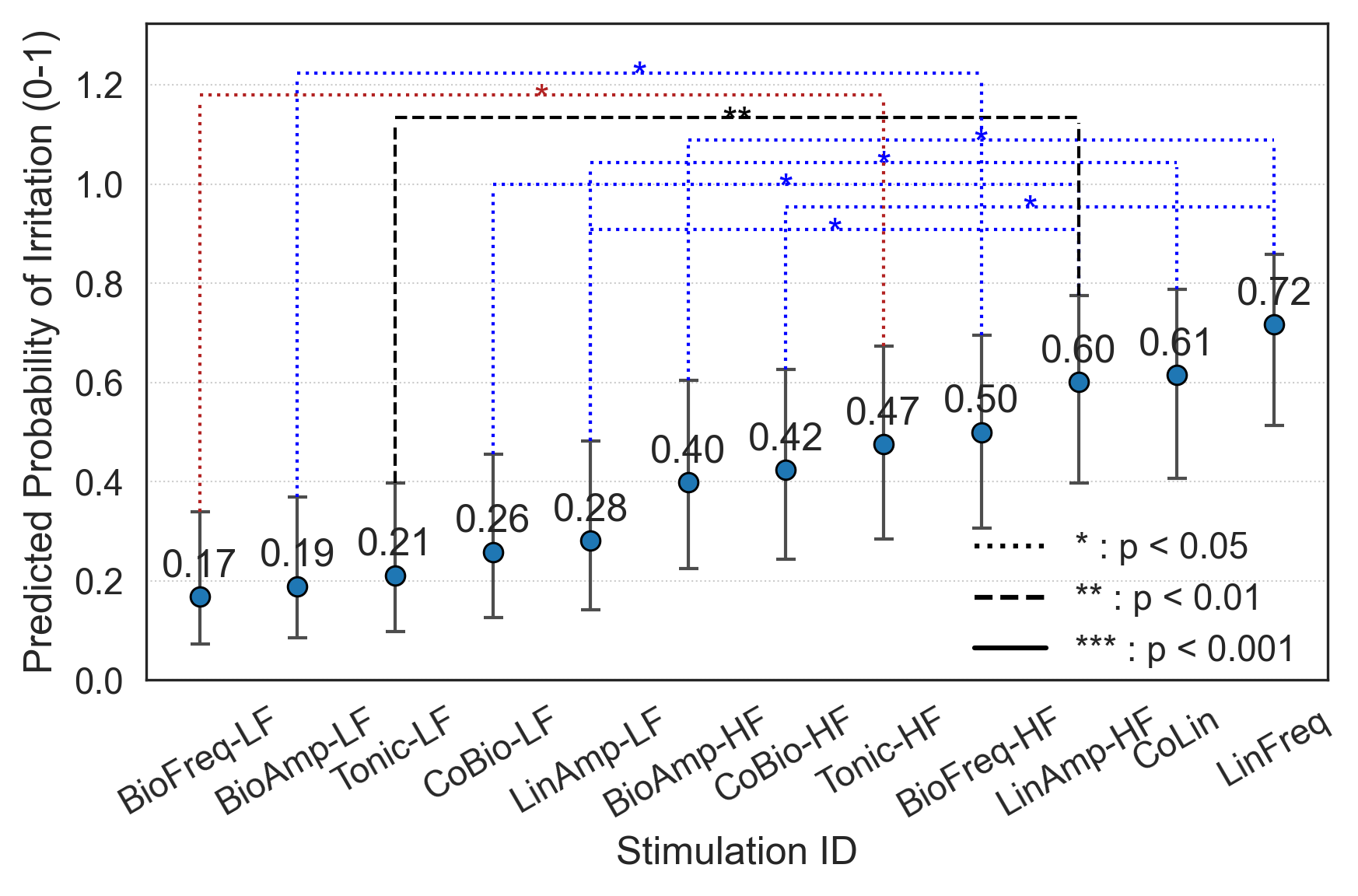}
    \label{fig::irritation_normal_probabilites}} 
    
    \subfloat[Video Condition]{\includegraphics[width=0.9\linewidth]{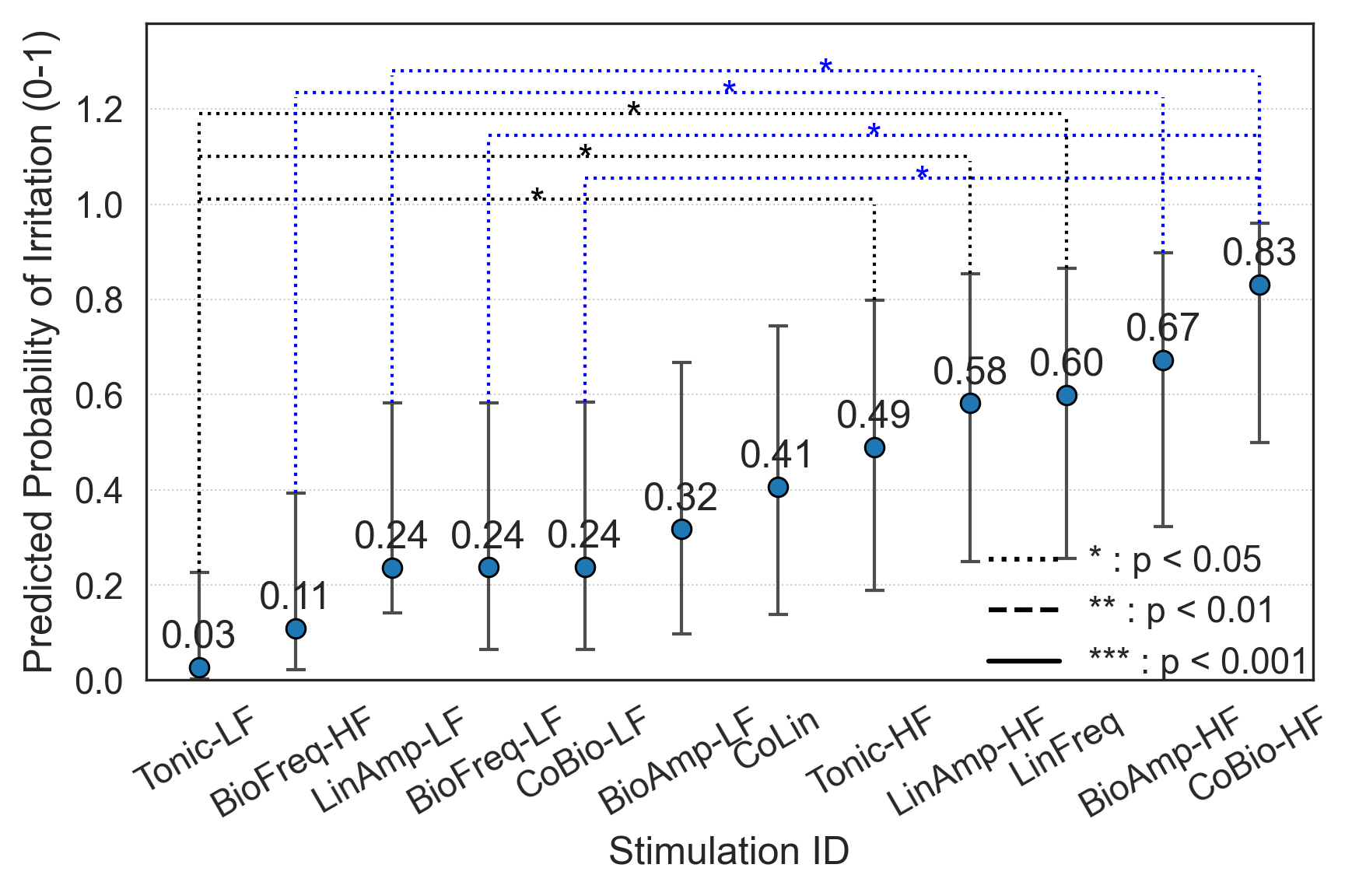}
    \label{fig::irritation_video_probabilites}} 
    
    \subfloat[Video+Movement Condition]{\includegraphics[width=0.9\linewidth]{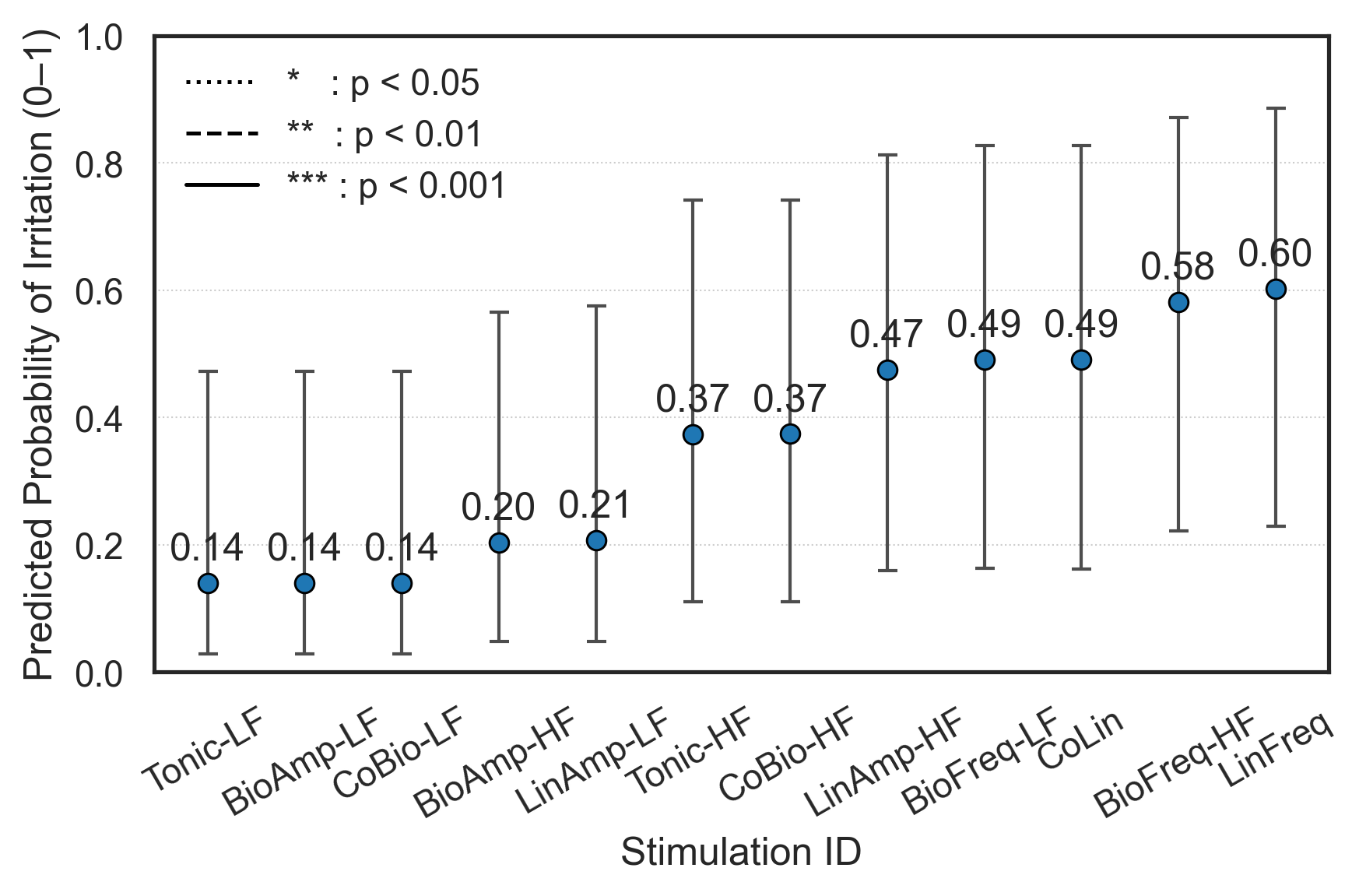}
    \label{fig::irritation_move_probabilites}} 
    
    \caption{The probability of producing irritation by each stimulation pattern under three different experimental conditions: a) Normal, b) Video, c) Video+Movement. Brackets with asterisks indicate the significant differences.}
    \label{fig::irritation_separate_condition}
\end{figure}

\Cref{fig::irritation_separate_condition} shows the predicted probability of reporting irritation for each stimulation pattern across the three experimental conditions. Similar to \Cref{fig::pressure_separate_condition} some redundant significance markers are omitted to maintain visual clarity.

In the Normal condition (\Cref{fig::irritation_normal_probabilites}), lower-frequency stimulation patterns consistently exhibited lower predicted probabilities of irritation, suggesting that stimulation frequency influences irritation more strongly than the specific modulation pattern. Pattern-wise, relative to Tonic-LF (which showed a comparatively low irritation probability), the linearly modulated patterns LinAmp-HF, CoLin, and LinFreq yielded significantly higher irritation probabilities. No other patterns differed significantly from Tonic-LF. Using BioFreq-LF (the lowest-probability pattern) as the reference, Tonic-HF and all patterns with higher predicted irritation than Tonic-HF differed significantly. Notably, BioAmp and CoBio did not differ from BioFreq-LF, even in their high-frequency variants, indicating that these patterns were effective at limiting irritation relative to other high-frequency stimulations.

The Video condition showed a different profile from the Normal condition. Both CoBio-HF and BioAmp-HF exhibited high irritation probabilities and were significantly higher than Tonic-LF, whereas BioFreq-HF shifted towards the lower-probability group.

In the Video+Movement condition, BioFreq-LF shifted towards higher irritation probabilities, whereas BioAmp-HF remained among the lower-probability patterns, consistent with the Normal condition.

\subsection{Top Stimulations based on Pressure-Irritation Trade-off}

In line with the study objectives, the optimal stimulation patterns would combine a high probability of eliciting pressure with a low probability of eliciting irritation. However, the preceding analyses showed that patterns associated with higher pressure often also produced higher irritation. We therefore define the top patterns as those that best balance the pressure-irritation trade-off.

\begin{figure}[t]
\centering{\includegraphics[width=0.95\linewidth]{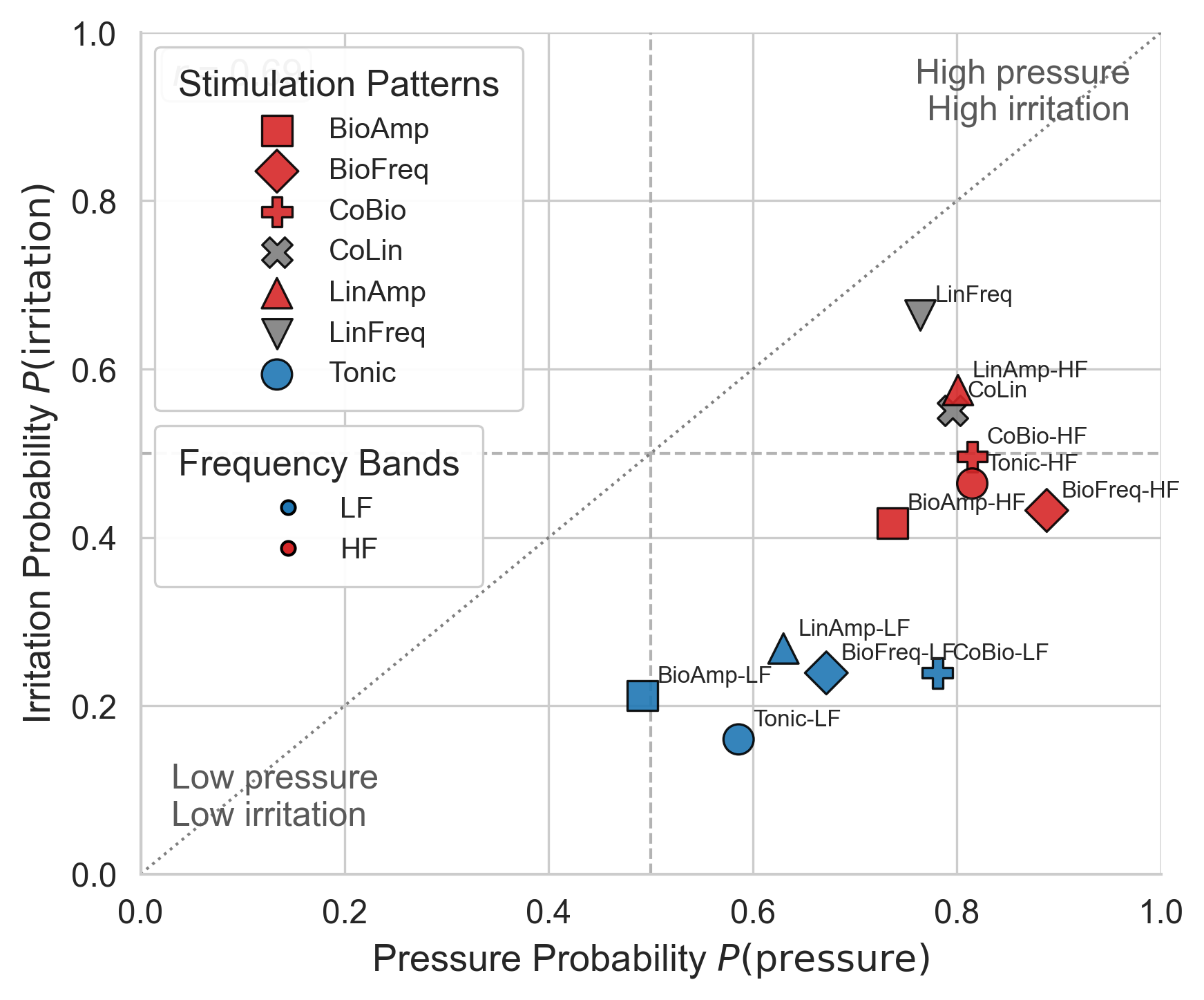}}
\caption{Stimulation patterns on the pressure-irritation probabilities plane.}
\label{fig::pressure_irritation_plane}
\end{figure}

To identify candidates, we considered the condition-specific results reported above. In addition, we fitted a pooled model to the full dataset with additive effects of stimulation pattern and experimental condition (i.e., \textit{Stimulation} + \textit{Condition}). Overall, effects of stimulation pattern on pressure and irritation were then obtained by averaging post hoc contrasts across conditions. As a sensitivity check, we also fitted a reduced model with stimulation pattern as the only fixed effect; this yielded the same conclusions. The predicted probabilities from the pooled model were used to position stimulation patterns in a pressure-irritation plane (\Cref{fig::pressure_irritation_plane}).

As shown in \Cref{fig::pressure_irritation_plane}, BioFreq-HF and CoBio-LF occupy favourable regions of the pressure-irritation plane, combining relatively high pressure probabilities with relatively low irritation probabilities; both are bio-inspired patterns.

Overall, CoBio-LF provided the most favourable balance. In the pooled pressure model, CoBio-LF did not differ from BioFreq-HF ($p>0.05$), which had the highest predicted pressure probability. And in the pooled irritation model, CoBio-LF did not differ from Tonic-LF ($p>0.05$), the lowest-irritation pattern. These results suggest that CoBio-LF can be considered a robust and comparatively conservative choice. Consistently, the condition-specific analyses (\Cref{fig::pressure_separate_condition,fig::irritation_separate_condition}) show that CoBio-LF remained competitive and did not differ from the best-performing patterns for pressure and irritation within each condition.

If some additional irritation can be tolerated to maximise pressure, BioFreq-HF is a suitable alternative. It achieved the highest predicted pressure probability in the pooled analysis, with moderate irritation (\Cref{fig::pressure_irritation_plane}) that was lower than LinFreq (the most irritating pattern) in the pooled model. Condition-specific results showed that BioFreq-HF produced the highest pressure under Normal and Video and ranked second under Video+Movement (\Cref{fig::pressure_separate_condition}). For irritation, BioFreq-HF did not differ from Tonic-LF in the Normal condition ($p>0.05$), but was higher than BioFreq-LF (the lowest-irritation pattern); it ranked among the lower-irritation patterns in Video, but shifted towards higher irritation under Video+Movement (\Cref{fig::irritation_separate_condition}).

Taken together, CoBio-LF is the safer overall choice, whereas BioFreq-HF offers higher pressure with a greater risk of irritation, particularly when combined with movement cues.

Condition-specific selection yielded minor differences. BioFreq-HF and CoBio-LF remained the strongest candidates in the Normal condition, and BioFreq-HF was again most favourable in Video. In Video+Movement, CoBio-HF replaced BioFreq-HF as the strongest high-pressure option, while CoBio-LF and Tonic-LF offered comparatively low irritation. Despite these differences, especially under Video+Movement, the pooled results are likely to be more robust given the limited number of participants and trials in the Video and Video+Movement conditions alone. More targeted data collection within a single condition would be required for definitive condition-specific pattern selection.

\subsection{Impacts of Visual and Movement Cues on Pressure and Irritation}

\begin{figure}[t]
\centering{\includegraphics[width=0.95\linewidth]{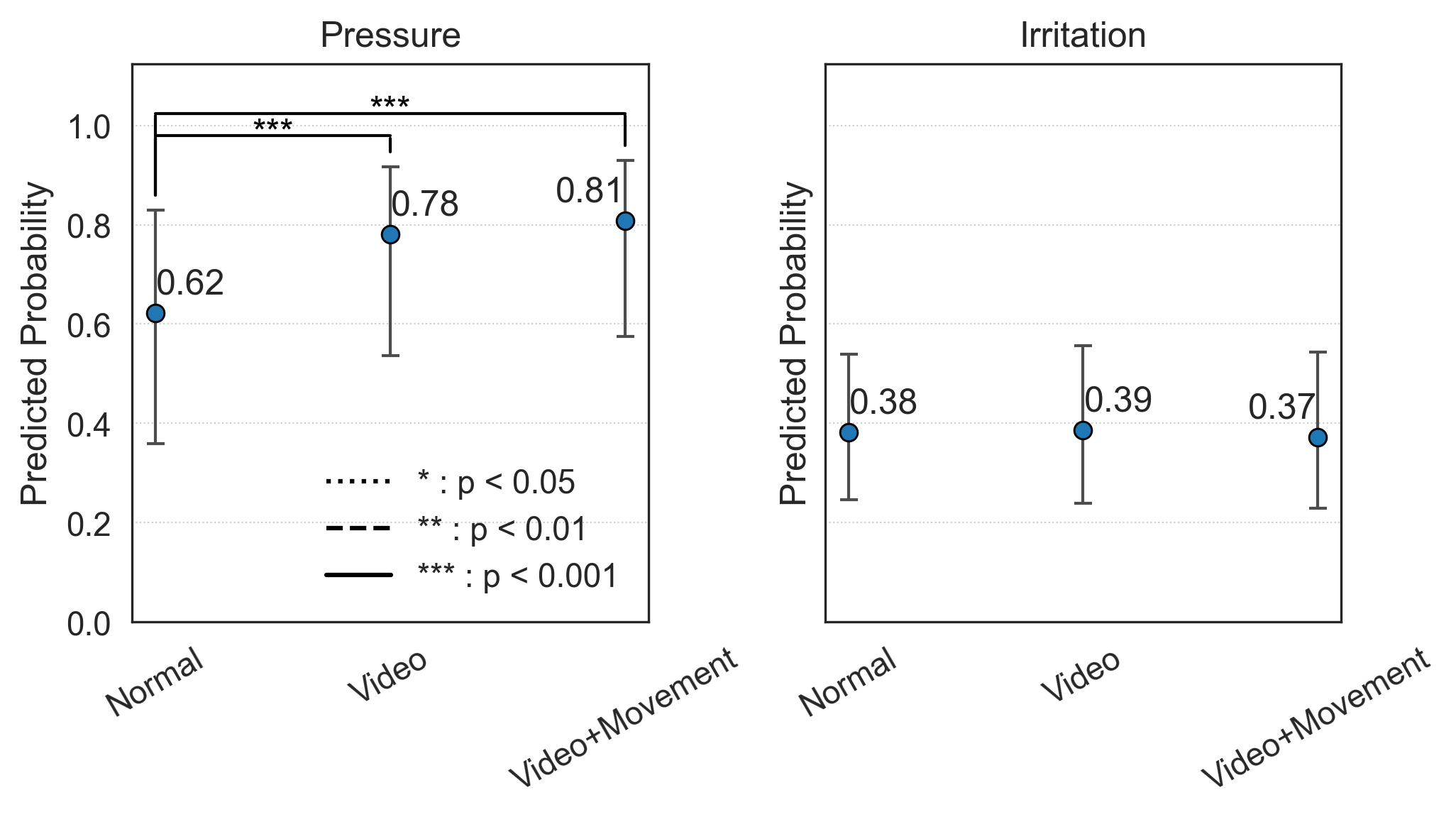}}
\caption{Effects of experimental conditions on perceived sensations, averaged across stimulation patterns.}
\label{fig::condition_effects_sensation}
\end{figure}

To assess the effects of visual and motion cues on pressure and irritation, we used the pooled GLMM fitted to the full dataset with additive effects of stimulation pattern and condition (i.e., \textit{Stimulation} + \textit{Condition}). Estimated marginal means were obtained by averaging over stimulation patterns, and post hoc contrasts between conditions are reported in \Cref{fig::condition_effects_sensation}.

As shown in \Cref{fig::condition_effects_sensation}, both Video and Video+Movement significantly increased the predicted probability of reporting pressure relative to Normal (left panel; $p<0.001$). Although Video+Movement yielded higher probabilities than Video, the difference was not statistically significant, suggesting no additive benefit of movement cues. For irritation, the absence of significant condition effects indicates that neither visual nor movement cues universally mitigate electrotactile discomfort, implying that improvements are likely driven by signal optimisation rather than cues alone (\Cref{fig::condition_effects_sensation}).

Decomposing the aggregate effects of conditions reveals that the efficacy of multimodal cues is highly dependent on the stimulation signal (\Cref{fig::pressure_separate_condition,fig::irritation_separate_condition}). While Video and Video+Movement increased pressure probability for most patterns, benefits were not uniform. Notably, LinAmp-HF performance decreased in both cue conditions, and Tonic-HF decreased under Video. The strongest pressure gains were observed for CoBio-HF, CoBio-LF, and BioAmp-HF (Video), and CoBio-HF, Tonic-LF, and CoLin (Video+Movement).

Irritation outcomes were similarly divergent. While cues reduced irritation for some patterns (e.g., BioFreq-HF and Tonic-LF under Video), they increased it for others. For instance, CoBio-HF showed a marked increase in irritation under Video, while BioFreq-LF and BioFreq-HF increased under Video+Movement. Crucially, CoBio-LF emerged as a standout pattern, combining substantial pressure gains with a concurrent reduction in irritation, distinguishing it as a robust candidate for realistic, multimodal applications.

\subsection{Impacts of Stimulation Patterns on Naturalness Score}
The second metric used to evaluate sensation quality was naturalness. Naturalness (ordinal) scores were modelled using a cumulative link mixed model (CLMM) with stimulation patterns as fixed effect and participant as a random intercept. Four models were fitted: three condition-specific models (Normal, Video, and Video+Movement) and one model using the full dataset. 

\begin{figure}[t]
\centering{\includegraphics[width=\linewidth]{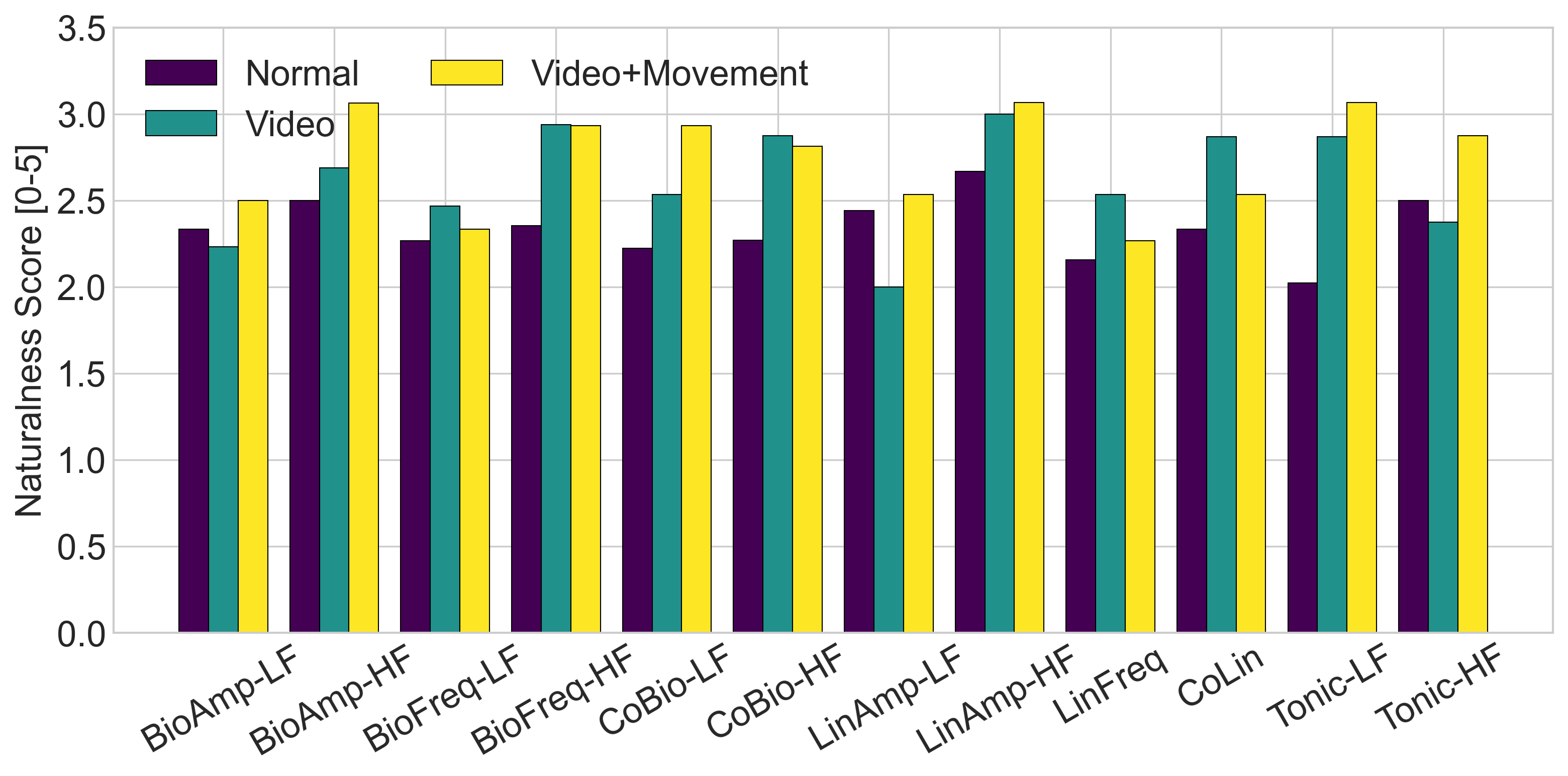}}
\caption{Mean naturalness scores for each stimulation pattern in different experimental conditions.}
\label{fig::naturalness_in_all_conditions}
\end{figure}

LRT revealed no significant stimulation pattern effects in the Normal or Video+Movement conditions. While an overall effect was observed in the Video condition, post hoc comparisons against the Tonic-LF as the reference failed to identify specific differences. Consequently, we restrict our analysis to descriptive trends for exploratory purposes (\Cref{fig::naturalness_in_all_conditions}).

\Cref{fig::naturalness_in_all_conditions} summarises the mean naturalness scores for each stimulation pattern across the three experimental conditions. Descriptively, LinAmp-HF consistently achieved the highest mean naturalness, and CoBio-LF and BioFreq-HF, the patterns identified as favourable in the pressure-irritation analysis, also achieved comparatively high naturalness scores. However, scores were tightly clustered across all patterns. This lack of separation suggests that a single ordinal naturalness rating may have been too imprecise to capture the multidimensional nature of the tactile experience across varying scenarios.

\subsection{Impacts of Visual and Movement Cues on Naturalness}
A CLMM was fitted to naturalness scores with experimental condition as the fixed effect and participant as a random intercept. The likelihood-ratio test indicated a significant effect of condition ($\chi^{2}(2)=18.09$, $p<0.001$). Post hoc contrasts showed that Video significantly increased naturalness relative to Normal ($p<0.01$), and Video+Movement also increased naturalness relative to Normal ($p<0.001$). In contrast, there was no evidence of a difference between Video and Video+Movement ($p>0.5$).

To examine how these cues affected individual stimulation patterns relative to Normal, we refer to \Cref{fig::naturalness_in_all_conditions}. Tonic-LF exhibited the largest relative gains under both Video and Video+Movement, followed by secondary improvements for BioFreq-HF and CoBio-HF under Video and CoBio-LF under Video+Movement. Thus, while LinAmp-HF maintained the highest absolute naturalness, Tonic-LF benefited most from the addition of multimodal cues.

\section{Discussion}\label{sec::discussion}
Previous work on bio-inspired electrotactile patterns often omitting amplitude modulation. Because amplitude is closely tied to perceived intensity and depends on individual thresholds, it has typically been ignored or treated via simple linear scaling rather than a bio-inspired formulation \cite{valle2024biomimetic, bayat2025effects, valle2018biomimetic}. Our framework, including the proposed estimation of $A_2$ and $B_2$ in \cref{eq::scaling_and_bias_formula}, enables the principled incorporation of amplitude modulation into the bio-inspired model.

Overall, co-modulating amplitude and frequency (CoBio-LF) yielded the most favourable pressure-irritation trade-off, increasing pressure while keeping irritation comparatively low. This aligns with intra-neural work showing that combining bio-inspired frequency modulation with linear amplitude modulation can enhance naturalness and tactile sensitivity in amputees \cite{valle2018biomimetic}. Naturalness did not differ significantly between patterns; however, exploratory summaries suggested higher mean naturalness for LinAmp-HF and BioAmp-HF, consistent with evidence that linearly increasing amplitude can enhance perceived softness \cite{yem2018effect}.

Adding video and movement cues significantly increased the probability of perceiving pressure, whereas no clear condition effect was observed for irritation. Irritation increased for some patterns under Video, with a small mitigating trend when movement was added. A plausible explanation is that video strengthens expectations (making unwanted sensations more salient), while movement divides attention between the scene and the motor task, reducing sustained focus on irritation. Both cues increased naturalness relative to Normal. Importantly, cue effects were pattern-dependent: although CoBio-HF and CoBio-LF both showed substantial pressure gains with video, only CoBio-LF concurrently reduced irritation, making it the more attractive option overall; Tonic-LF also showed favourable changes, improving irritation. These results underline the need to evaluate electrotactile patterns in realistic, multimodal contexts.

\paragraph{Framework limitations and future work}
The framework assumes linear relationships between frequency and firing rate, and between amplitude and activation rate within each time bin; future work could test non-linear alternatives (e.g., logarithmic/exponential), consistent with evidence that recruited fibre counts can increase exponentially with current amplitude during intra-neural stimulation \cite{bourbeau2011computational}. 

Pulse width was fixed to control the number of effective parameters, preserve independence in the firing and activation rates formulations, and simplify calibration; future studies could integrate pulse width into the framework, or substitute it with amplitude, to better characterise its role. 

For electrode design derived from afferent responses, we prioritised temporal patterning over fixed spatial grouping; alternative aggregation strategies may yield further gains. For example, dynamically varying spatial activation showed to enhance perceived softness \cite{suga2023softness}; future work could develop bio-inspired spatio-temporal variants and compare them against both the present approach and \cite{suga2023softness}.

\paragraph{Experimental and metric limitations}
We evaluated sensation type and naturalness, but the single-item Likert naturalness rating showed limited sensitivity: scores clustered between 2 and 3, reducing separation between patterns and statistical power. 
Moreover, "naturalness" appeared participant dependent (comfort vs realism/similarity to the VR scenario), and these interpretations shifted when video and movement were introduced. Future work should consider multidimensional scales or composite naturalness metrics to disentangle these facets \cite{anwar2023factors}. 

To keep the experiment duration manageable given three conditions and 12 patterns, each pattern was presented once per participant in the Video and Video+Movement sessions. This reduced sampling may have limited statistical power, which could partly explain the absence of significant pattern effects for pressure in the Video condition and the non-significant likelihood-ratio tests for pressure and irritation in the Video+Movement condition. Future work should prioritise deeper data collection within a single condition to enable more reliable condition-specific pattern comparisons.

In addition, testing many stimulation patterns reduced post hoc power because of the large number of simultaneous contrasts. The present statistical and exploratory results can therefore be used to shortlist a smaller set of candidate patterns for focused follow-up studies aimed at identifying the best-performing stimulation.

Because the Video and Video+Movement sessions were always conducted last, learning or habituation may partly confound the estimated cue effects. Although naturalness and pressure tended to increase with trial order, the clear step-change between conditions is consistent with a genuine condition effect. Nonetheless, future studies should mitigate ordering effects by randomising the sequence of conditions and/or using repeated sessions.
 
\paragraph{Movement and task scope}
No participant reported discomfort or fading attributable to hand movement; perceived intensity remained consistent, in line with \cite{yao2022encoding}. This suggests that, for the limited movements tested, movement-related changes may lie below a critical threshold for stable perception and skin-electrode contact. However, movement complexity and duration were constrained, so these results do not generalise to more dynamic or prolonged interactions; further work should test richer movements and longer use. Finally, focusing on a single scenario (button pressing) limits generalisability, and broader tactile interactions are needed for more comprehensive validation. 

\paragraph{Touch model considerations}
TouchSim was selected for its accuracy, complexity, and reliability, but its computational demands may limit speed for complex stimuli; faster or simplified variants may be preferable for real-time systems \cite{ouyang2020simplified}. Importantly, replacing the touch model does not invalidate the proposed framework; the framework is therefore adaptable across touch models without compromising the underlying approach.

\section{Conclusions}\label{sec::conclusion}
In this study, we developed a bio-inspired framework that maps touch-simulator responses to temporally varying electrotactile stimulation parameters. The results indicate that bio-inspired modulation of amplitude and frequency, particularly their co-modulation, can improve electrotactile feedback by increasing pressure perception while maintaining irritation at moderate levels. We also showed that visual and movement cues exert significant, pattern-dependent effects, highlighting the need to evaluate electrotactile feedback in realistic, multimodal scenarios. Future work should address the current limitations of the framework and assess its generalisation to broader tasks and applications to further improve sensation quality and practical usability.

\section{Acknowledgment}
\noindent The authors sincerely thank 6GIC, for their support and all who contributed to the realisation of this work, including the volunteers who generously participated in the experiment.

\IEEEtriggeratref{15}

\bibliographystyle{IEEEtran}
\bibliography{References.bib}

\end{document}